\documentclass[preprint2]{emulateapj}

\newcommand{\fer}{{\it Fermi}}
\newcommand{\bzcat}{{\it Roma-BZCAT}}
\newcommand{\wse}{{\it WISE}}
\usepackage{graphicx}
\usepackage{longtable}
\usepackage{xcolor}

\slugcomment{version \today: fm}

\shorttitle{The infrared-$\gamma$-ray connection}
\shortauthors{F. Massaro \& R. D'Abrusco}

\begin{document}
\title{The infrared--gamma-ray connection.\\ A \wse\ view of the extragalactic gamma-ray sky}
\author{F. Massaro\altaffilmark{1,2,3} \& R. D'Abrusco\altaffilmark{4}}

\affil{Dipartimento di Fisica, Universit\`a degli Studi di Torino, via Pietro Giuria 1, I-10125 Torino, Italy}
\affil{Istituto Nazionale di Fisica Nucleare, Sezione di Torino, I-10125 Torino, Italy}
\affil{Istituto Nazionale di Astrofisica - Osservatorio Astrofisico di Torino, via Osservatorio 20, I-10025 Pino Torinese, Italy}
\affil{Smithsonian Astrophysical Observatory, 60 Garden Street, Cambridge, MA 02138, USA}

\begin{abstract}

{ Using data from the \wse\ all-sky survey we discovered that the non-thermal infrared (IR) emission of blazars, the largest known population 
of extragalactic $\gamma$-ray sources, has peculiar spectral properties. In this work, we confirm and strengthen our previous 
analyses using the latest available releases of both the \wse\ and the \fer\ source catalogs. We also show that there is a tight correlation between 
the mid-IR colors and the $\gamma$-ray spectral index of \fer\ blazars. We name this correlation {\it the infrared--$\gamma$-ray connection}. 
We discuss how this connection links both the emitted powers and the spectral shapes of particles accelerated in jets arising from blazars 
over ten decades in energy. Based on this evidence, we argue that {\it the infrared--$\gamma$-ray connection} is stronger than the well
 known {\it radio--$\gamma$-ray connection}.}

\end{abstract}


\section{Introduction}
\label{sec:intro}

Blazars are one of the most extreme class of radio-loud active galaxies whose emission extends from radio to TeV energies. They generally 
show extreme variability at all wavelengths and with timescales spanning from weeks to minutes, evidence of superluminal motions, high 
and variable polarization, flat radio spectra \citep[see e.g.][]{urry95}, recently observed even below 
$\sim$1GHz \citep[i.e. ,][]{ugs3,massaro13d} and a characteristic double bumped spectral energy distribution \citep[SED; see also][for a 
recent review]{massaro09}. 

Since the launch of the \fer\ satellite \citep{atwood09}, blazars { have been identified} as the dominant class of $\gamma$-ray sources, 
not only extragalactic. { Blazars account for} about 1/3 of the \fer\ detected objects \citep{acero15} and { likely for} a significant fraction 
{ of} the unidentified/unassociated $\gamma$-ray sources \citep[UGSs;][]{massaro12a,ugs2}. Together with star forming 
{ regions} \citep[e.g.][]{ackermann12a} and radio galaxies \citep[e.g.][]{dimauro14,lobes}, blazars produce a significant contribution 
to the extragalactic $\gamma$-ray background \citep[][and references therein]{ajello12,ajello14,review}. 

At optical frequencies blazars are historically { split} in two subclasses: BL Lac objects and flat spectrum radio quasars. The former, 
{ that will be indicated in this paper} as BZBs { following} the nomenclature { introduced by} the \bzcat\ \citep{massaro15b}, 
show featureless spectra and/or with weak absorption lines of equivalent width lower than 5$\AA$, while the latter, { usually indicated} as 
BZQs, have quasar-like optical spectra \citep{stickel91,falomo14}.

{ Since} 2010, { the NASA} Wide-field Infrared Survey Explorer \citep[\wse;][]{wright10} { has} mapped the sky in the infrared 
(IR) at 3.4, 4.6, 12, and 22 $\mu$m, { making possible the investigation} of the mid-IR properties of a large, statistically significant, 
sample of confirmed blazars. We discovered that \fer\ blazars inhabit a region of the mid-IR color-color diagram, built with the \wse\ 
magnitudes, well separated from the location of other extragalactic sources \citep{paper1,paper2}. This 2-dimensional region in the mid-IR 
color-color diagram [3.4]-[4.6]-[12] $\mu$m was originally indicated as the {\it \wse\ Gamma-ray Strip}, and it is the projection of a 
3-dimensional volume in the [3.4]-[4.6]-[12]-[22] $\mu$m mid-IR color space known as 
the \wse\ {\it locus} of $\gamma$-ray blazars \citep{ugs1,wibrals}. 

{ These findings} led to the development of different procedures to search for $\gamma$-ray blazar candidates within the positional 
uncertainty regions of the \fer\ UGSs { that selected hundreds of IR sources as candidate blazars}. Thanks to an extensive optical 
spectroscopic follow-up campaign \citep{paggi14,massaro14,refined} we confirmed the { nature of} hundreds of new $\gamma$-ray
 blazars \citep[see also][]{massaro15c,landoni15,ricci15} { and assessed the reliability of our association methods}. 
 Optical { spectroscopy was also used} to determine the nature of \fer\ sources classified as active galaxies of uncertain type 
 \citep[AGUs;][]{nolan12,ackermann11} and/or blazar candidates of uncertain type \citep[BCUs;][]{acero15,ackermann15a}.

Here we present an update of the {\it \wse\ Gamma-ray Strip} obtained { by} combining the latest releases of both the \wse\ All-Sky 
catalog\footnote{http://wise2.ipac.caltech.edu/docs/release/allsky/} and the Fermi Large Area Telescope Third Source 
Catalog \citep[3FGL;][]{acero15}. Additionally, for the first time, we discuss the link found between the \wse\ mid-IR colors and the 
$\gamma$-ray photon index, comparing it with the radio--$\gamma$-ray connection \citep[e.g.][]{ackermann11b}.

For our numerical results, we use cgs units unless stated otherwise. Gamma-ray photon index, $\Gamma$ is defined by the usual convention 
on the flux density, $N(E)\propto\,E^{-\Gamma}$, being $N(E)$ the number of $\gamma$-ray photons detected per unit of time, area and 
energy. \wse\ magnitudes are in the Vega system and are not corrected for the Galactic extinction since, as shown in our previous analyses, 
such correction affects significantly only the magnitude at 3.4$\mu$ for sources lying at low Galactic latitudes~\citep[see e.g.][]{wibrals}. 
{ \wse\ bands are indicated as $w1$, $w2$, $w3$ and $w4$ and corresponds to the following nominal wavelengths: 
3.4$\mu$m, 4.6$\mu$m, 12$\mu$m and 22$\mu$m, respectively}.

\section{Sample Selection}
\label{sec:samples}

{ The analysis presented hereby has been carried out using three distinct samples of sources extracted from the \bzcat\
catalog of confirmed blazars, the 3FGL catalog of \fer\ sources and catalog of radio sources extracted from the NVSS and SUMSS surveys, 
respectively. Then, we searched the mid-IR counterparts in the \wse\ all-sky catalog for all sources in each of these samples (the fraction of
objects in each sample with a \wse\ association are reported  
reported in Table~\ref{tab:detect}). It is worth noting that the Two Micron All Sky 
Survey (2MASS)~ \citep{skrutskie06} counterparts of the \wse\ sources are automatically reported in the \wse\ catalog.}

The first sample includes all BZBs and BZQs listed in the latest version of the \bzcat\ \citep[i.e., v5.0; ][]{massaro15b} and associated to 
a \fer\ source { belonging to} the Third Catalog of Active Galactic Nuclei Detected by the \fer\ Large Area Telescope 
\citep[3LAC][]{ackermann15a}. This sample is the { update of the sample} used to build the {\it \wse\ Gamma-ray Strip} in~\cite{paper2}, 
{ and contains} 1036 $\gamma$-ray blazars: 610 BZBs and 426 BZQs. Adopting the statistical criterion described in our previous analyses 
we { identified} as blazar counterpart 
the closest IR source within 3\arcsec.3 from the position reported in the \bzcat\ \citep[e.g.,][]{wibrals}. This criterion was chosen because 
source positions of the \bzcat, even if often more precise than those reported in the 3LAC, are taken from various catalogs with different 
positional uncertainties. Given the high detection rate of \fer\ blazars in the first 3 \wse\ filters we only focus on the analysis in the 
{ color diagram} built with the 3.4$\mu$m, 4.6$\mu$m and 12$\mu$m magnitudes.

The second sample { is composed by} BCUs of the 3FGL with a \wse\ counterpart within 10\arcsec\ . This is the typical radio 
positional uncertainty used in the 3LAC, { usually corresponding} to that of the { radio} NRAO VLA Sky Survey \citep[NVSS][]{condon98} 
and/or the Sydney University Molonglo Sky Survey \citep[SUMSS;][]{mauch03} used to search for counterparts of \fer\ sources. 

The third sample includes all the radio sources of both the NVSS and the SUMSS catalog lying within the positional uncertainty region 
of the UGSs at 95\% level of confidence and with a \wse\ counterpart detected at least at 3.4$\mu$m, 4.6$\mu$m and 12$\mu$m, 
within a maximum radius obtained by { adding the radio to the mid-IR positional uncertainties at 1$\sigma$ level of confidence.}

\begin{table} 
	\caption{Number of sources detected at IR frequencies in both the \wse\ and the 2MASS catalogs and for { the three} samples 
	considered in our analysis.}
	\label{tab:detect}
	\begin{tabular}{|crrrr|rrr|r|}
	\hline
	Sample  & $w1$ & $w2$ & $w3$ & $w4$ & J & H & K & Total \\
	\hline 
	\noalign{\smallskip}
	BZB  & 603 & 603 & 595 & 497 & 521 & 523 & 528 & 610 \\
	BZQ  & 419 & 419 & 419 & 405 & 207 & 206 & 215 & 426 \\
	BZB $\cup$ BZQ & 1022 & 1022 & 1014 & 902 & 728 & 729 & 743 & 1036 \\
	\hline 
	\noalign{\smallskip}
	BCU & 569 & 569 & 553 & 433 & 319 & 328 & 325 & 571 \\
	\hline 
	\noalign{\smallskip}
	UGS & 1546 & 1465 & 558 & 315 & 580 & 607 & 584 & 1548 \\
	\noalign{\smallskip}
	\hline
	\end{tabular}\\
	Note:  \wse\ bands are indicated as $w1$, $w2$, $w3$ and $w4$ while 2MASS magnitudes are $J$, $H$, $K$.
\end{table}

\section{Gamma-ray blazars at IR wavelengths}
\label{sec:gir}

\begin{figure*}[!ht]
	\includegraphics[height=8.cm,width=9.cm,angle=0]{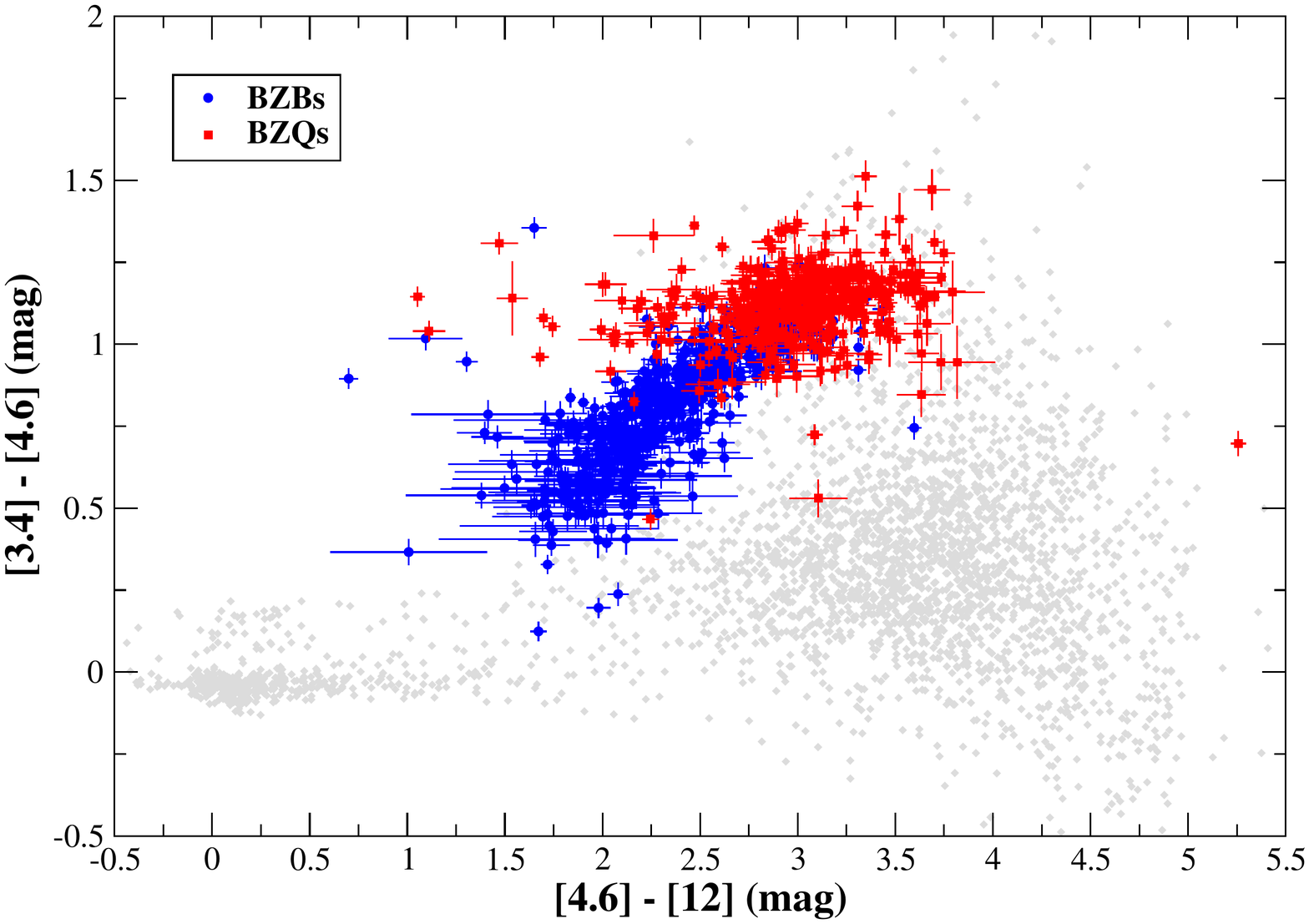}
	\includegraphics[height=8.cm,width=9.cm,angle=0]{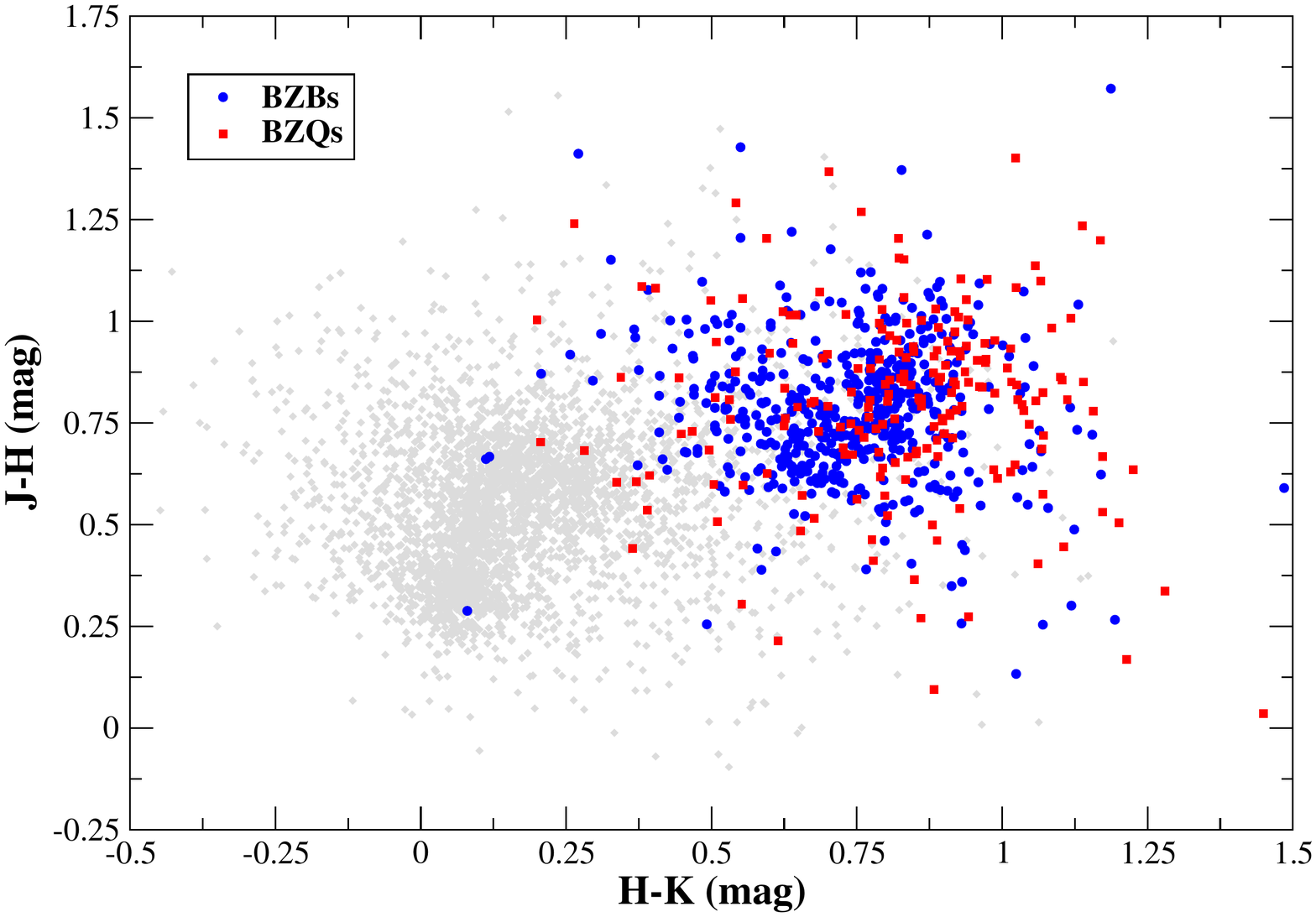}
	\caption{Left) { Distribution of sources belonging to the {\it WISE Gamma-ray Strip} (see \S~\ref{sec:samples}), in the 
	$w2-w3$ vs $w1-w2$ color-color diagram. BZBs and BZQ are displayed as blue circles 
	and red squares respectively. $\sim$3000 generic IR sources (grey circles) selected at high Galactic 
	latitudes are also show}. The { separation} between the mid-IR colors of the \fer\ blazars and that of other sources 
	randomly chosen appears evident. Right) { Distribution of the same sample in the near-IR 2MASS $H-K$ vs $J-H$ 
	color-color diagram. The distinction between the region occupied by 
	the \fer\ blazars and other celestial objects appears clear also in this case but there is { no} neat separation between 
	BZBs and BZQs.}}
	\label{fig:strip}
\end{figure*}

The extremely high detection rate of \fer\ blazars at mid-IR frequencies ($\approx$99\%) at 3.4 and 4.6 $\mu$m 
(see Table~\ref{tab:detect}), strongly supports the use of statistical methods combined with optical spectroscopic 
campaigns toward the identification of \wse\ selected blazar-like sources { that can be responsible for} the 
UGSs \citep{paggi13,acero13}.

{ The distribution of the updated sample of \fer\ blazars in the WISE color-color diagram shown in Fig.~\ref{fig:strip}, confirms 
our previous results \citep[see e.g.][]{paper1} with a 5 times larger sample}. The use of \wse\ colors as distinct markers 
of the non-thermal emission of blazars is clearly supported by these findings.

For comparison, in the color-color diagram generated using the J,H and K magnitudes in the 2MASS catalog 
(see Fig.~\ref{fig:strip}), although a clear separation between generic IR sources and the region preferentially populated by 
\fer\ blazars { is visible}, no { peculiar} color pattern is observed for BZBs and BZQs. Moreover,
a large fraction of the \fer\ BZQs considered do not have a 2MASS counterparts (see Table~\ref{tab:detect}), strongly limiting the 
{ potential} efficiency and completeness of methods { for the association of UGS} based on 2MASS photometry. 
The region occupied by the \fer\ blazars in the 2MASS color-color plot is similar to that previously indicated by Chen et al. (2005) 
for a different sample of BZBs and BZQs. However, about 30\% of the blazars in the Chen et al. (2005) sample show 
near-IR colors corresponding to a blackbody spectral shape, { suggesting} a possible contributions from their host galaxies. 
Such strong contamination { is not present} at mid-IR wavelengths.

\section{The infrared--gamma-ray connection}
\label{sec:connection}

\begin{figure*}[!ht]
	\includegraphics[height=8.cm,width=9.cm,angle=0]{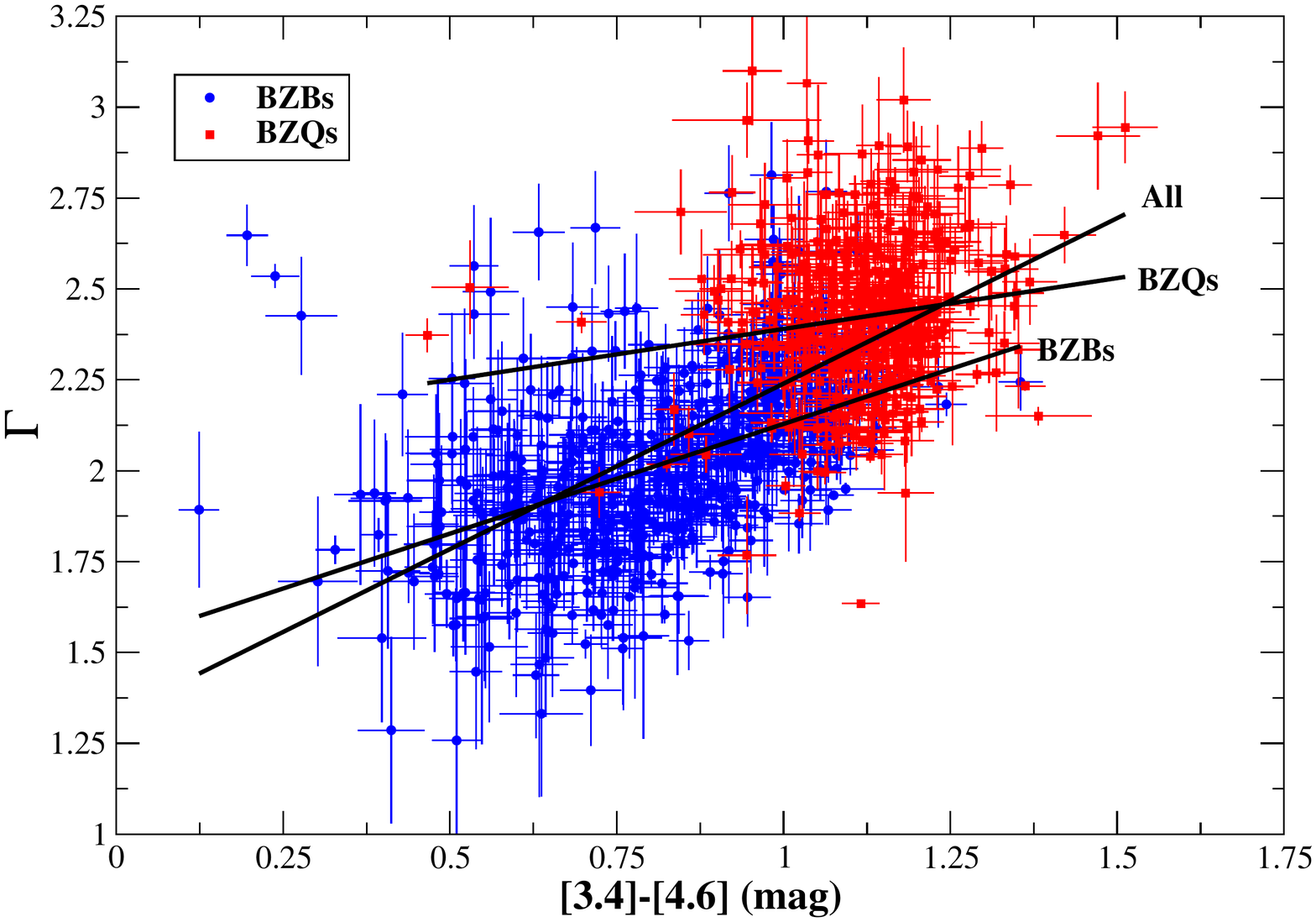}
	\includegraphics[height=8.cm,width=9.cm,angle=0]{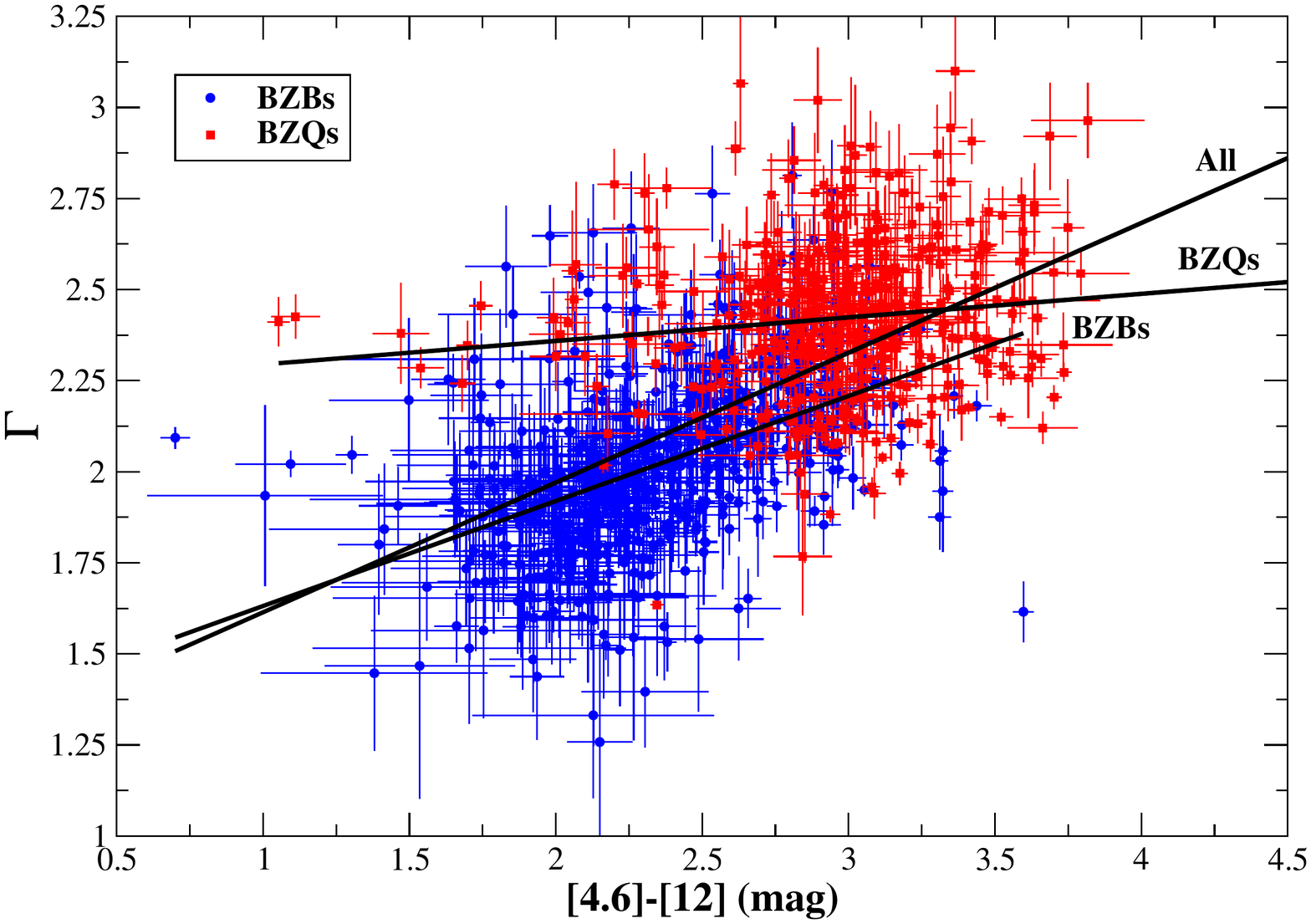}
	\includegraphics[height=8.cm,width=9.cm,angle=0]{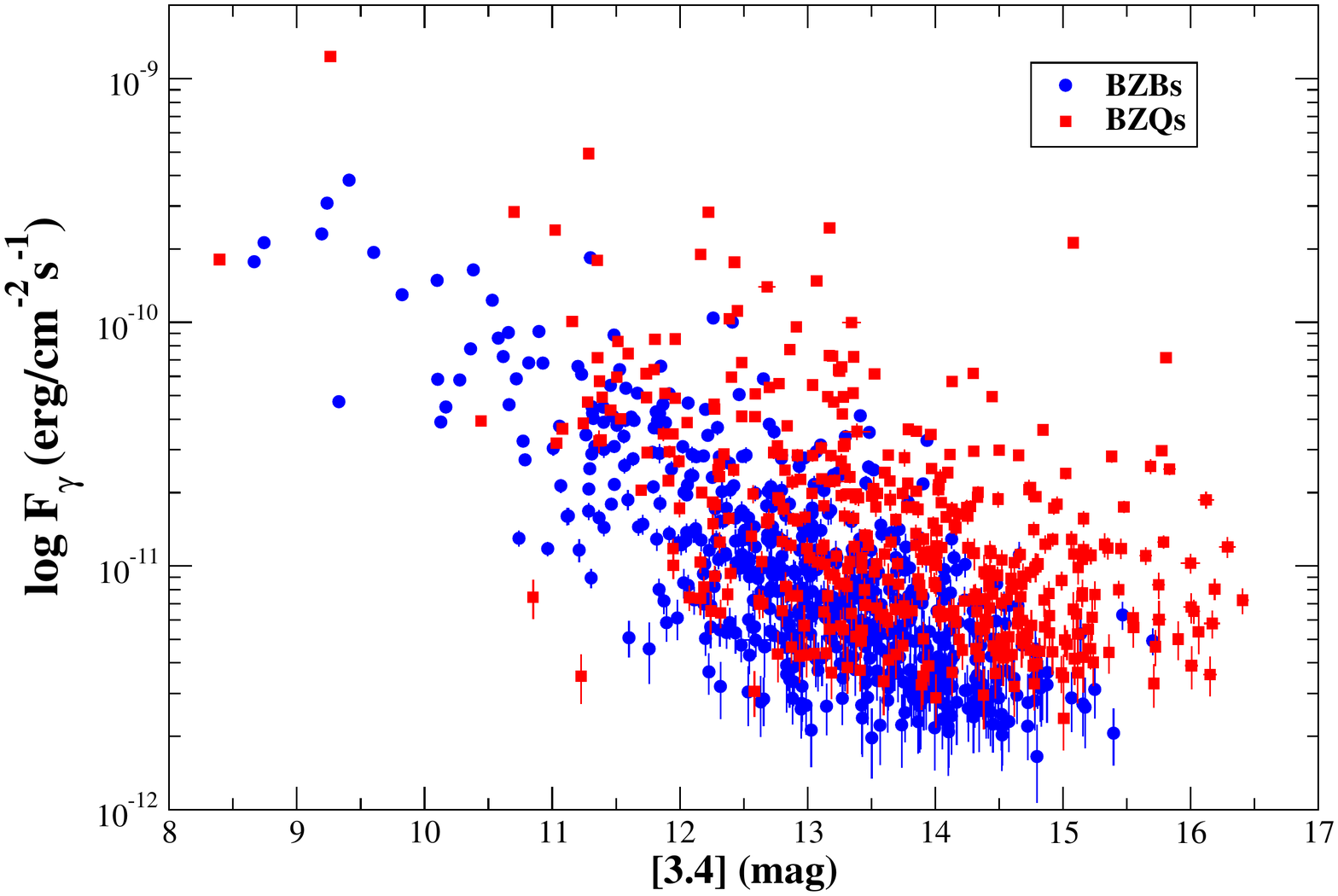}
	\includegraphics[height=8.cm,width=9.cm,angle=0]{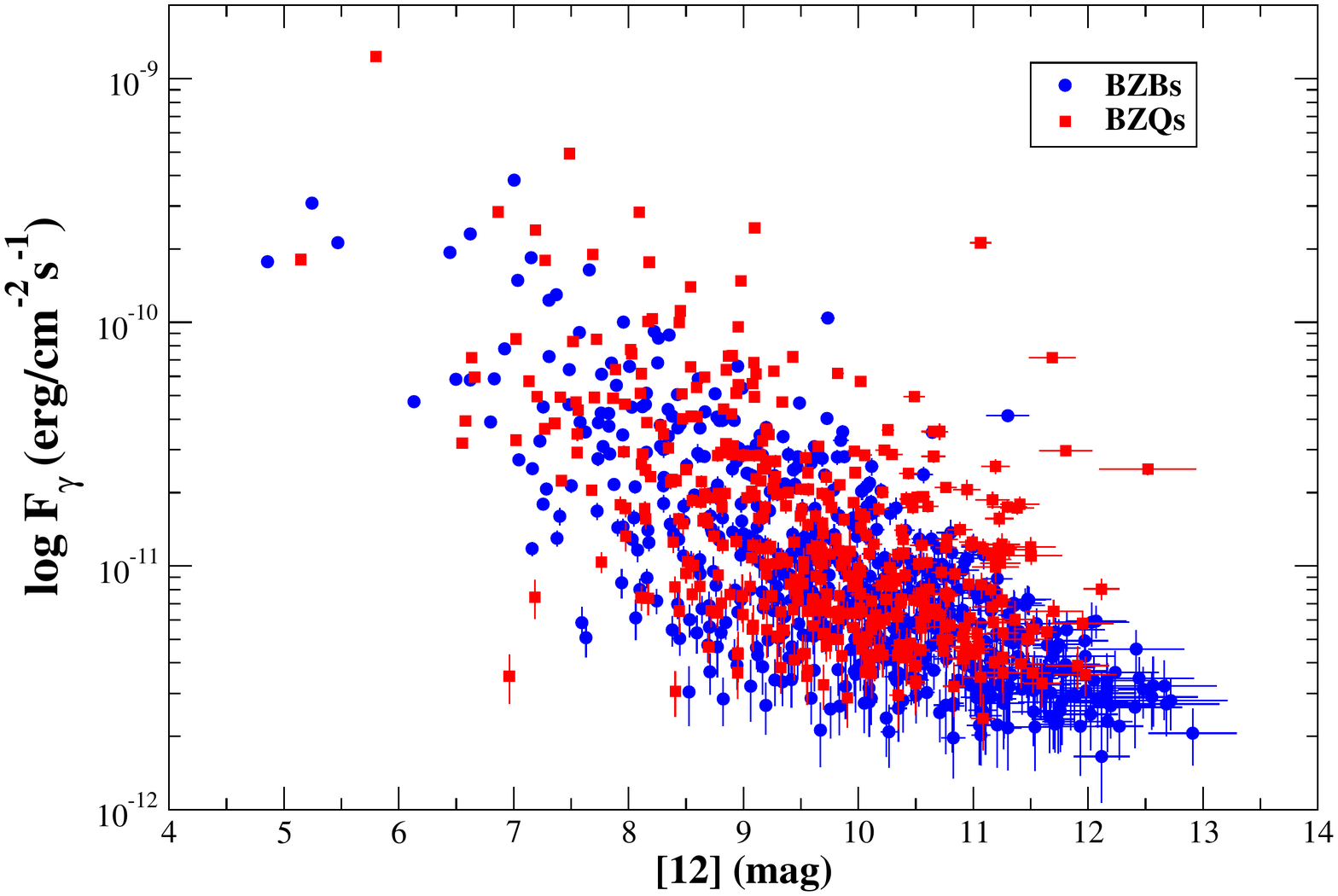}
	\caption{Upper panels): the $\Gamma$ $vs$ [3.4]-[4.6] $\mu$m (left) and the $\Gamma$ $vs$ [4.6]-[12] $\mu$m (right) diagrams {
	of the {\it WISE Gamma-ray Strip} sample}.
	BZBs are shown in blue circles while BZQs in red squares. The correlations between the $\gamma$-ray spectral shape and the \wse\ mid-IR 
	colors are highlighted by the regression lines (black) { for the whole sample and for} BZB and BZQ { subclasses separately}. 
	Lower panels): { distribution of the {\it WISE Gamma-ray Strip} sources in the wse\ magnitudes at [3.4] (left) and [12] (right) 
	$\mu$m vs $\gamma$-ray flux density planes. The distribution on the right shows that 12 $\mu$m emission from BZBs 
	is less contaminated by the host galaxy than at 3.4 $\mu$m}.}
	\label{fig:corr}
\end{figure*}

{ To assess the significance of the {\it infrared--$\gamma$-ray connection}}, we performed a statistical analysis of the 
correlation between the $\gamma$-ray photon index $\Gamma$ of the \fer\ blazars and the 
\wse\ mid-IR colors, since colors are a surrogate of the spectral index. We found a strong correlation between $\Gamma$ and both 
the \wse\ colors [3.4]-[4.6] $\mu$m and [4.6]-[12] $\mu$m as shown in Fig.~\ref{fig:corr}. This result constitutes the first direct 
evidence of the existence of the {\it IR--$\gamma$-ray connection}. 

In both cases, the correlation is more statistically significant for BZBs rather than for BZQs. The linear correlation coefficient $\rho$ 
{ estimated for the} $\Gamma$ vs [3.4]-[4.6] $\mu$m { correlation} is 0.47 { for} BZBs, with negligible probability to be spurious.
On the other hand, for the BZQs, we obtained $\rho=$0.15 calculated with 419 sources and still with a negligible chance probability, 
given the large number of objects considered. When using the data for the whole { sample} of 1022 \fer\ blazars, the linear correlation 
coefficient, { unsurprisingly}, increases to $\rho=$0.65 { thanks to the quite distinct} distributions of both BZBs and BZQs
in the $\Gamma$ $vs$ [3.4]-[4.6] $\mu$m plane.

A similar situation { is observed} in the $\Gamma$ $vs$ [4.6]-[12] $\mu$m { plane}, where the correlation { coefficients for the 
595 BZBs and the 419 BZQs separately are $\rho=$0.48 and $\rho=$0.13 respectively}, both with negligible chance probability. The 
correlation { coefficient} for the whole \fer\ blazar sample (1014 sources) is $\rho$=0.60, lower than the previous one due to the { scatter 
in the plane introduced by the} larger uncertainties on the mid-IR magnitudes at 12$\mu$m. 
The slope of the best fit regression lines in both cases, as expected, is driven by the stronger correlations found for the BZBs. 

Finally, we did not find any significant trend between the 2MASS J, H and K colors, corrected for Galactic extinction, 
and the \fer\ $\gamma$-ray spectral index $\Gamma$.

\section{Comparison with the radio--gamma-ray connection}
\label{sec:comparison}

Since the release of the first observations carried out with the {\it Compton Gamma-ray Observatory}, a link between radio and 
$\gamma$-ray emissions of blazars appeared evident \citep[e.g.,][]{stecker93}. This well established {\it radio--$\gamma$-ray connection} 
{ has mostly driven the searches} for 
counterparts associable with $\gamma$-ray sources \citep{mattox97}, and { has been invoked to justify} radio-follow up observations 
of the positional uncertainty regions for UGSs to date \citep[e.g.,][and references therein]{schinzel15}. This connection between the blazar 
emission at $\sim$ 1GHz and at GeV energies is due to the correlation between their flux densities and, consequently, their powers. 

Thanks to the large increase of the number of blazars detected by \fer\, it has become possible to study the presence of biases and 
selection effects that have to be taken into account to confirm the radio--$\gamma$-ray connection. Intrinsic variability, redshift dependence, 
source misidentifications and incorrect associations \citep[see e.g.][]{mucke97}, to name a few effects that can mimic a spurious trend, 
were not found to affect this { correlation}. Based on this analysis, the existence of the radio--$\gamma$-ray connection \citep[see also][and 
references therein]{mahony10} was recently confirmed by Ackermann et al. (2011b). 

{ For the sake of comparison, it is worth stressing that the {\it radio--$\gamma$-ray connection} is based on the correlation 
between the radio and $\gamma$-ray flux densities. A similar trend also exists for the mid-IR emission as shown in the lower 
panels of Fig.~\ref{fig:corr}, which display the 3.4 $\mu$m and the 12$\mu$m magnitudes vs $\gamma$-ray flux density for 
both confirmed \fer\ BZBs and BZQs.}

{ Based on the discussion in the previous Section, it appears evident that the {\it IR--$\gamma$-ray connection} is stronger 
than the radio--$\gamma$-ray one, because it is not only a link between flux densities but it also involves a tight connection between 
spectral shapes of the emitting particles.}

\section{The uncertain gamma-ray sky}
\label{sec:ugs}

We stress the strength of the {\it IR--$\gamma$-ray connection} by showing that potential counterparts of { 3FGL} UGSs 
selected via their WISE Mid-IR colors are in good agreement with the correlation between the [3.4] and [12] WISE colors and 
the $\gamma$-ray $\Gamma$ spectral index discussed in \S~\ref{sec:connection}.

We considered all the \wse\ sources with a radio counterpart that lie within the positional uncertainty regions of the UGSs at 95\% level 
of confidence. By applying the Kernel Density Estimation (KDE) technique \citep[see e.g. ][and reference therein]{dabrusco09}, as in our 
previous analyses \citep[see e.g.,][]{paper1,massaro13c,refined}, we estimated the probability density function of the IR color distribution for 
the \fer\ blazar population and we selected as blazar-like candidates those { located} within the 95\% isodensity contours drawn from the 
KDE probabilities in the [3.4]-[4.6]-[12] $\mu$m diagnostic diagram.

\begin{figure}
	\includegraphics[height=8.cm,width=9.cm,angle=0]{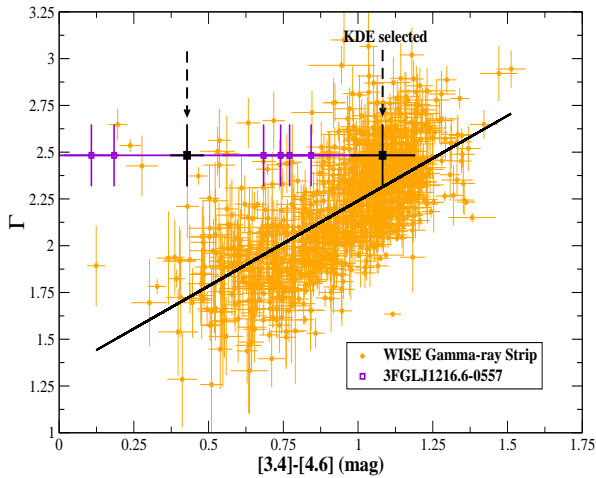}
	\caption{The $\Gamma$ $vs$ [3.4]-[4.6] $\mu$m { distribution of WISE sources with a radio counterpart 
	within the region of positional uncertainty at 95\% level of confidence for} the 
	UGS 3FGL J1216.6-0557 (squares). Orange circles in the background are { confirmed} \fer\ 
	blazars { and} the black line { shows the} regression between their spectral parameters. The { arrows indicate} the 
	only two \wse\ sources detected at [3.4]-[4.6]-[12] $\mu$m for which it was possible to compute the KDE probability. { The only 
	\wse\ source selected with the KDE method is the closest to the regression line correspondent 
	to the {\it IR--$\gamma$-ray connection}.}}
	\label{fig:single}
\end{figure}

In Fig.~\ref{fig:single} we show, as an example, the 8 \wse\ sources with a radio counterpart that lie within the positional uncertainty region 
at 95\% level of confidence of the UGS 3FGL J1216.6-0557. The { only} \wse\ source selected by the KDE analysis is the 
closest to the regression line { for} the { \wse\ [3.4]-[4.6] $\mu$m vs $\gamma$-ray spectral shape distribution of all the \fer\ blazars 
(BZBs and BZQs)} considered in our analysis (see \S~\ref{sec:samples} and \S~\ref{sec:connection} for more details).

\begin{figure*}
	\includegraphics[height=8.cm,width=9.cm,angle=0]{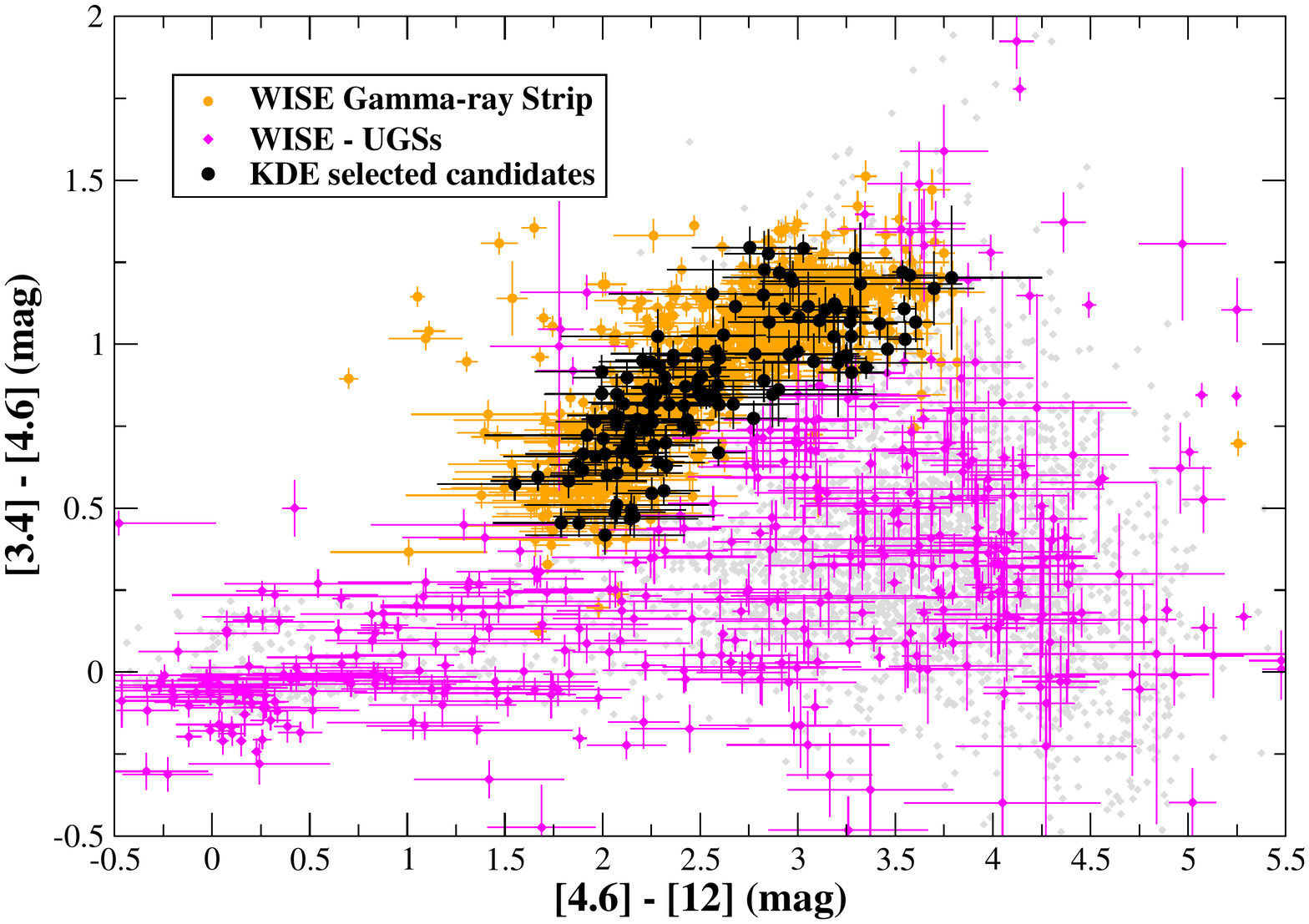}
	\includegraphics[height=8.cm,width=9.cm,angle=0]{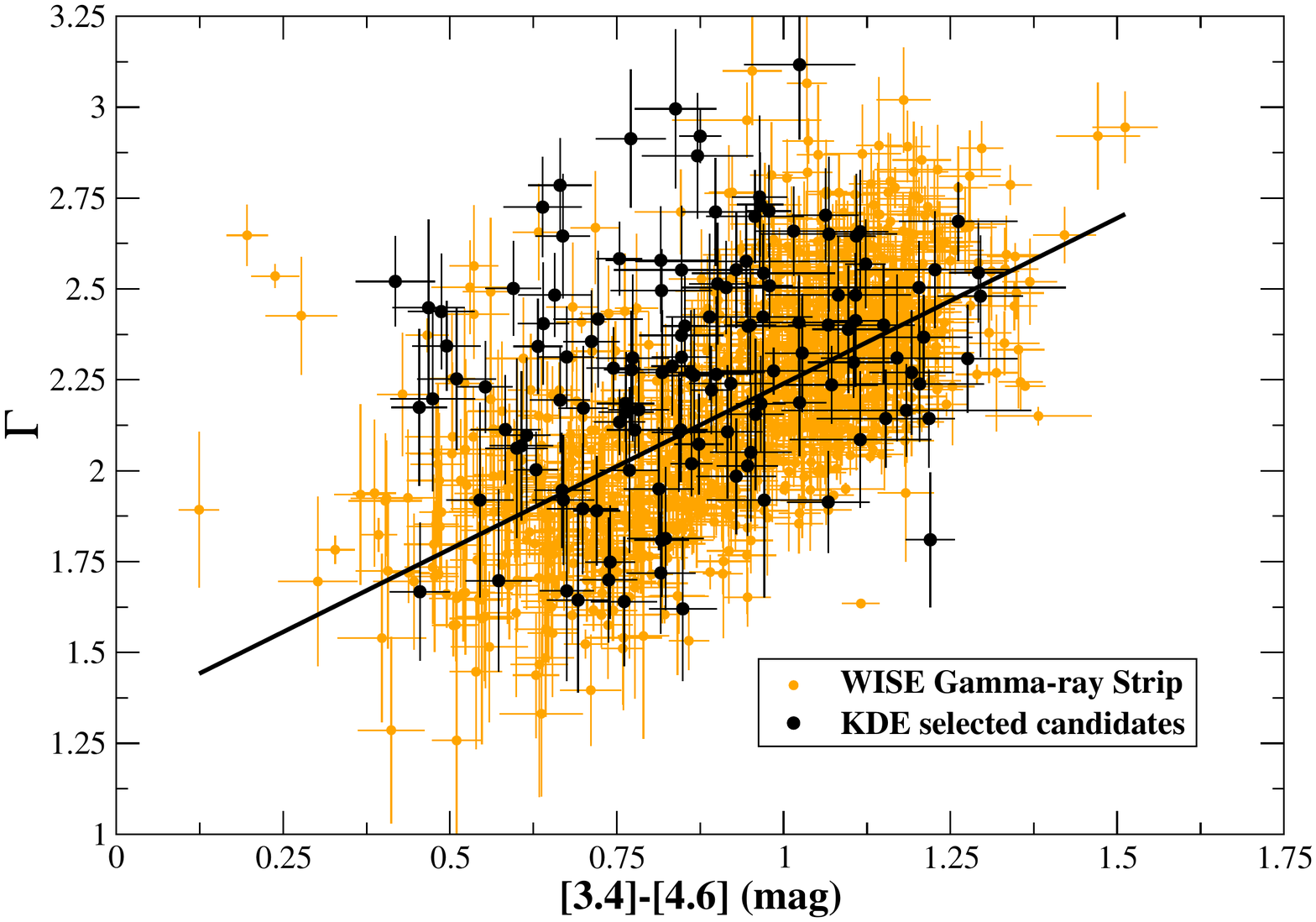}
	\includegraphics[height=8.cm,width=9.cm,angle=0]{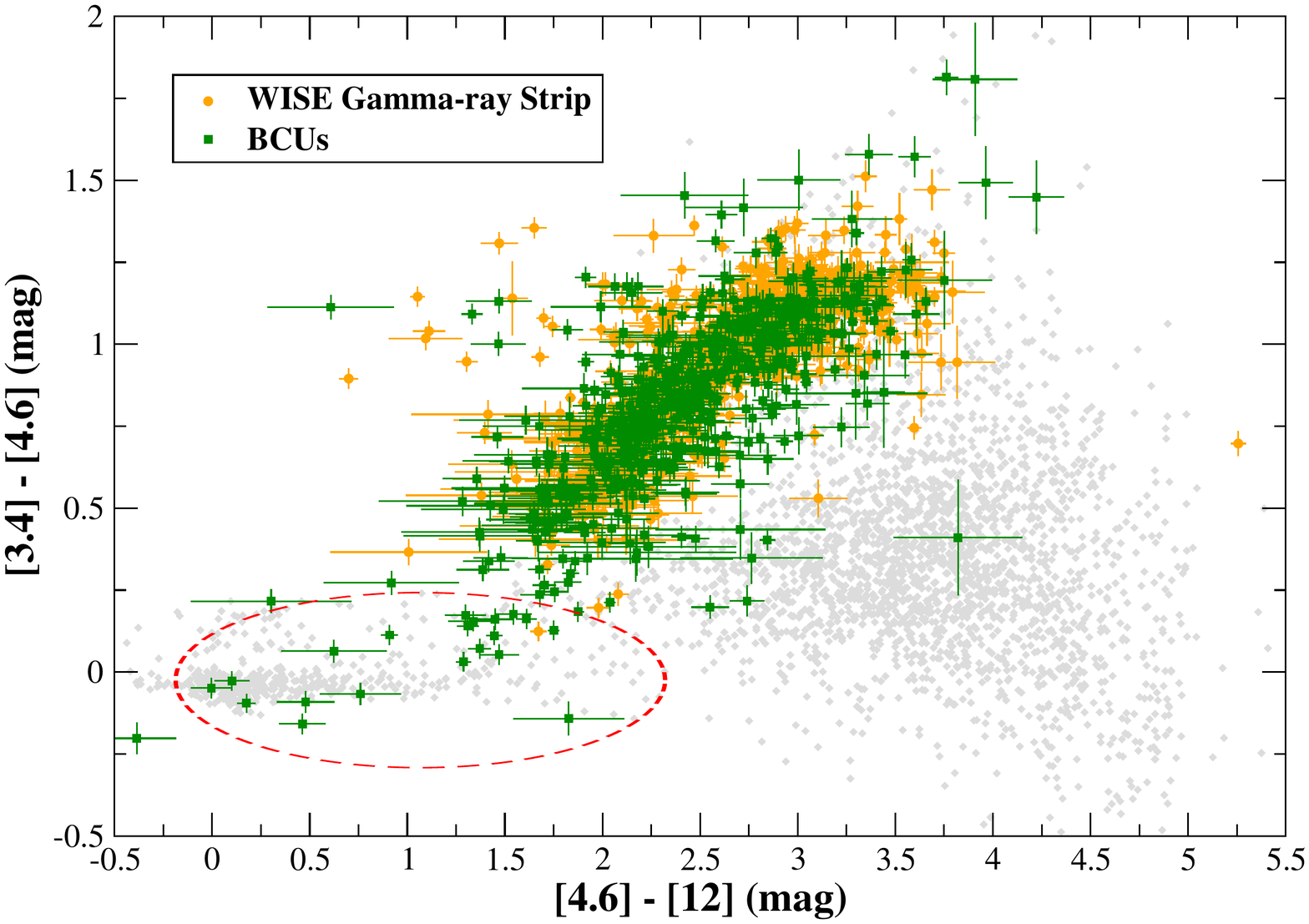}
	\includegraphics[height=8.cm,width=9.cm,angle=0]{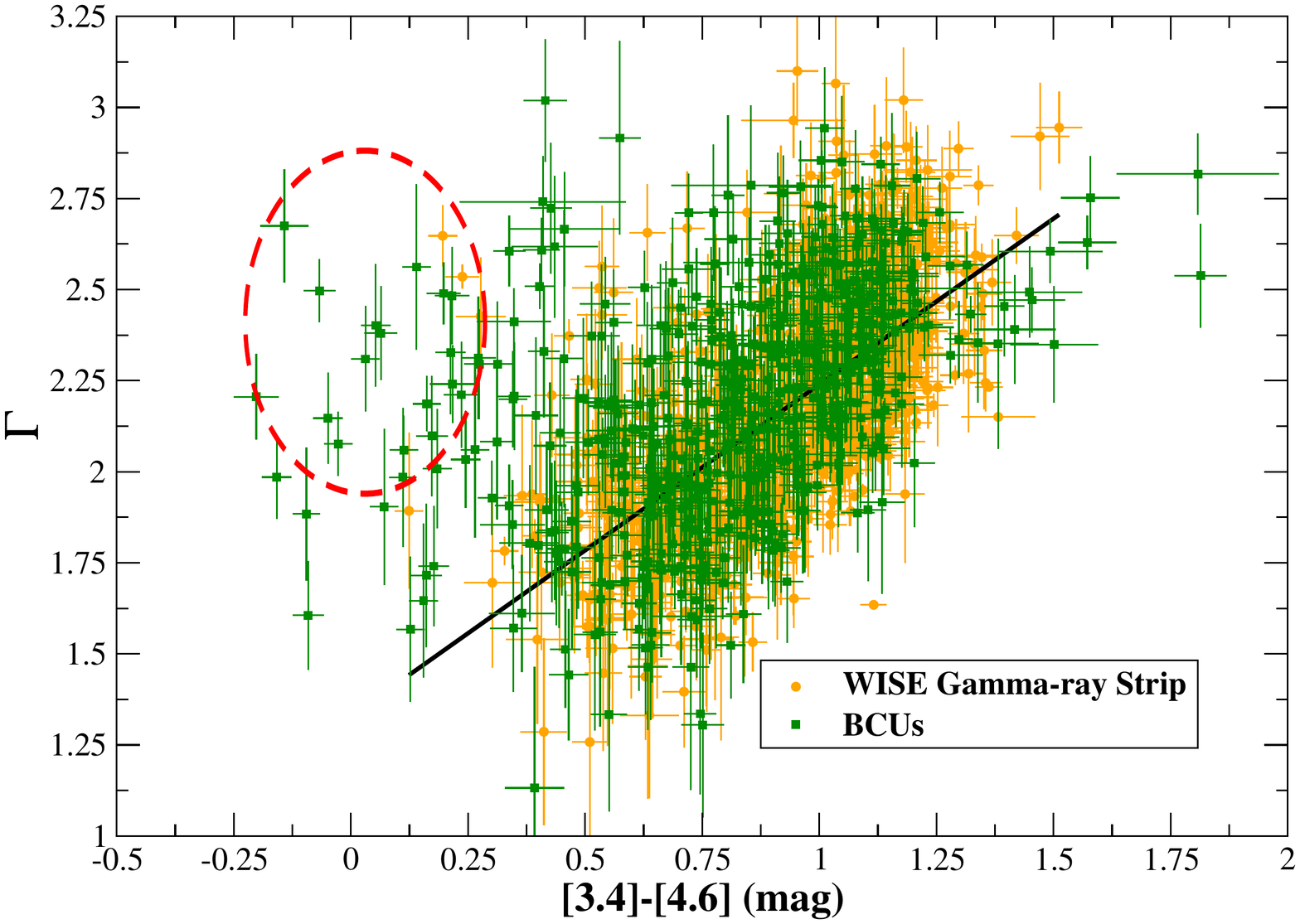}
	\caption{The [3.4]-[4.6]-[12] $\mu$m color-color { diagrams} (left panels) and the 
	$\Gamma$ $vs$ [3.4]-[4.6] $\mu$m diagrams (right panels) { for blazar-like sources
	associated to 3FGL UGSs (upper panels) and 3FGL BCUs (lower panels), respectively.}
	{ A large fraction of BCUs (i.e., more than 90\%) are}
	consistent with the { \wse\ colors of confirmed} \fer\ blazars, { i.e.} {\it \wse\ Gamma-ray Strip} { (lower left)}. 
	The dashed ellipse highlights a small { number} of BCUs 
	whose \wse\ counterparts have mid-IR colors similar to { those of} normal 
	galaxies and do not follow the correlation between the IR and the 
	gamma-ray spectral shape { (lower right)}. 
	In the { upper} left panel 
	we show the comparison between all the \wse\ sources lying within the positional uncertainty regions of the 3FGL UGSs with a radio 
	counterpart (magenta diamonds) and the sources of the {\it \wse\ Gamma-ray Strip}. { The} \wse\ sources { that 
	are selected according to the KDE method are shown as} black circles. These \wse\ sources are { mostly} 
	consistent with the {IR--$\gamma$-ray connection} { (upper right)}.
	In all panels \fer\ blazars are shown { as} orange circles { regardless of their spectral class}. Generic mid-IR sources 
	{ are shown in the background} as grey circles { in the two panels on the left}.}
	\label{fig:diag}
\end{figure*}

In general, by applying the same technique to { all the 987 UGSs listed in the 3FGL to date}, we selected 130 blazar-like potential 
counterparts out of 1548 \wse\ sources with a radio counterpart lying within the 95\% positional uncertainty regions. 
{ As shown in the $\Gamma$ $vs$ [3.4]-[4.6] $\mu$m diagram (Fig.~\ref{fig:diag}, top left)}, these candidates { are clearly 
located in} the region { occupied} by the \fer\ blazars, thus following the {\it IR-$\gamma$-ray connection}. It is { worthwhile to
stress} that such consistency it is not granted {\it a priori} since the \wse\ colors of \fer\ blazars simply { reflect the fact} that 
their mid-IR emission is dominated by non-thermal radiation independently { of} any $\gamma$-ray information. { As a consequence}, 
the {\it IR--gamma-ray connection} cannot be a consequence of the existence of {\wse\ $\gamma$-ray Strip}. 

To complete our investigation we also analyzed the 3FGL BCUs. As expected, { the mid-IR colors of a large fraction of them 
(i.e., more than 90\%) are} similar to { the \wse\ colors of confirmed} \fer\ blazars and follow the correlation between 
$\Gamma$ and [3.4]-[4.6] $\mu$m (\ref{fig:diag}) { that we have denominated the 
{\it IR--$\gamma$-ray connection}}. A large fraction of { these blazar-like sources} are also expected to be BZBs, 
in agreement with the recent results of several optical spectroscopic campaigns \citep[see e.g.][for more details]{shaw13,crespo16a,crespo16b} 
{ which find that more than 70\% of the observed sources are BZBs.}

\section{Summary and Discussion}
\label{sec:summary}

Five years after the discovery that the non-thermal emission of $\gamma$-ray blazars can be traced using mid-IR colors 
obtained { from the photometry of} the \wse\ all-sky survey, we present an updated analysis based on the latest releases of 
both the \wse\ and the \fer\ catalogs. Then, for the first time, we discuss on the existence of a {\it IR--$\gamma$-ray connection} 
for the \fer\ blazars that appears to be at least as strong as the well-known radio--$\gamma$-ray one. 

Our results can be summarized as follows.

\begin{enumerate}
	\item Using the largest sample of \fer\ blazars available to date, we confirmed that this extreme class of active galaxies { 
	occupied a narrow and well defined region} in the mid-IR 
	color-color { plane, the so called}  {\it \wse\ Gamma-ray Strip}.
	\item Comparing mid-IR \wse\ and near-IR 2MASS diagnostic diagrams, we { confirmed} that in the latter \fer\ blazar are 
	distinct from generic 2MASS sources, { even though} 
	no clear color-color trend appears as in the former. \fer\ blazars { have} a low detection rate in the 2MASS catalog, 
	mainly due the { 2MASS} higher flux limit { compared to \wse\ survey}. These reasons { advice against the use of}  
	near-IR colors to search for potential blazar-like counterparts of the UGSs. 
	\item We describe the { statistically significant} correlations between the $\gamma$-ray photon index and the mid-IR colors 
	{ for both the whole sample of \fer\ blazars and the BZBs and BZQs spectral classes separately}, the basis of the 
	{\it IR--$\gamma$-ray connection}. { This correlation} appears ``stronger'' than the well known radio--$\gamma$-ray 
	connection because { it} involves the spectral shapes 
	of the \fer\ blazars over $\sim$10 orders of magnitude and not only their flux densities. 
	\item { We argue that the peculiar mid-IR colors of \fer\ blazars do not depend on} their $\gamma$-ray 
	photon index. { In turn,} the {\it IR--$\gamma$-ray connection is} unexpected. { We have also highlighted} 
	its potential use to implement the search for blazar-like counterparts of the \fer\ UGSs.
	\item We show that a large fraction of the BCUs listed in the 3FGL whose \wse\ counterpart have mid-IR colors consistent 
	with the {\it \wse\ Gamma-ray Strip}, { have \fer\ spectral index values consistent} with the 
	{\it IR--$\gamma$-ray connection}.
\end{enumerate}

{ Taking advantage of the overwhelming fraction of \fer\ blazars detected in the \wse\ all-sky survey (i.e., $\sim$99\%) in the first two mid-IR filters, 
we suggest that a comprehensive investigation of their IR properties at the light of the {\it IR--$\gamma$-ray connection} 
described in this paper, can represent a powerful tool to reveal the real fraction of \fer\ blazars hidden within the sample of UGSs.}

\acknowledgements
We thank the anonymous referee for useful comments that led to improvements in the paper.
F.M. gratefully acknowledges the financial support of the Programma
Giovani Ricercatori -- Rita Levi Montalcini -- Rientro dei Cervelli (2012) awarded by the Italian Ministry of Education, Universities and Research (MIUR).
This research has made use of data obtained from the high-energy Astrophysics Science Archive
Research Center (HEASARC) provided by NASA's Goddard Space Flight Center.
The NASA/IPAC Extragalactic Database
(NED) operated by the Jet Propulsion Laboratory, California
Institute of Technology, under contract with the National Aeronautics and Space Administration.
Part of this work is based on the NVSS (NRAO VLA Sky Survey):
The National Radio Astronomy Observatory is operated by Associated Universities,
Inc., under contract with the National Science Foundation and on the VLA Low-frequency Sky Survey (VLSS).
The Molonglo Observatory site manager, Duncan Campbell-Wilson, and the staff, Jeff Webb, Michael White and John Barry, 
are responsible for the smooth operation of Molonglo Observatory Synthesis Telescope (MOST) and the day-to-day observing programme of SUMSS. 
The SUMSS survey is dedicated to Michael Large whose expertise and vision made the project possible. 
The MOST is operated by the School of Physics with the support of the Australian Research Council and the Science Foundation for Physics within the University of Sydney.
This publication makes use of data products from the Wide-field Infrared Survey Explorer, 
which is a joint project of the University of California, Los Angeles, and 
the Jet Propulsion Laboratory/California Institute of Technology, 
funded by the National Aeronautics and Space Administration.
This publication makes use of data products from the Two Micron All Sky Survey, which is a joint project of the University of 
Massachusetts and the Infrared Processing and Analysis Center/California Institute of Technology, funded by the National Aeronautics 
and Space Administration and the National Science Foundation.
TOPCAT\footnote{\underline{http://www.star.bris.ac.uk/$\sim$mbt/topcat/}} 
\citep{taylor05} for the preparation and manipulation of the tabular data and the images.


{}

\end{document}